\documentclass{cpbtex}
\usepackage{float}       %设置图片浮动位置的宏包
\usepackage{subfigure}   %插入多图时用子图显示的宏包
\usepackage{graphicx}  % Include figure files
\usepackage{caption}
\begin{document}

\title{Decoherence in the Pure Dephasing Spin-Boson Model with Hermitian or Non-Hermitian Bath}
% Title should be concise; avoid abbreviations if possible; and not begin with `A', `An', `The', or `Study on'.
\author{Yue-Hong Wu(吴月红)$^{1,2}$ and \ Ning-Hua Tong（同宁华）$^{1,2}$\thanks{Corresponding author. E-mail:~nhtong@ruc.edu.cn}\ \ \ \ \ \ \ \ \ \ \ \ \ \ \ \ \ \ \ \ \ \ \ \ \ \ \ \ \ \ \ \ \ \ \ \ \  \ \ \ \ \ \ \ \ \ \ \ \ \ \ \ \ \ \ \ \ \ \ \ \ \ \ \ \ \ \ \ \ \ \\
$^{1}${School of Physics, Renmin University of China, Beijing 100872, China}\ \ \ \ \ \ \ \ \ \ \ \ \ \ \ \ \ \ \ \ \ \ \ \ \ \ \ \ \ \ \ \ \ \ \ \ \ \ \ \ \ \ \ \ \ \ \ \ \ \ \ \ \ \ \ \ \ \ \ \ \ \ \ \ \ \ \ \ \ \ \ \ \ \ \ \\  % The line break was forced via \\
$^{2}${Key Laboratory of Quantum State Construction and Manipulation (Ministry of Education), \ \ \ \ \ \ \ \ \ \ \ \ \ \ \ \ \ \ \ \ \ \ \ \ \ \ \ \ \ \ \ \ \ \ \ \ \ \ \ \ \ \ \ \ \ \ \ \ \ \ \ \ \ \ \ \ \ \ \ \ \ \ \ }\\{Renmin University of China, Beijing 100872, China \ \ \ \ \ \ \ \ \  \ \ \ \ \ \ \ \ \  \ \ \ \ \ \ \ \ \  \ \ \ \ \ \ \ \ \  \ \ \ \ \ \ \ \ \  \ \ \ \ \ \ \ \ \  \ \ \ \ \ \ \ \ \ \ \ \ \ \ \ \ \ \ } \\ % The line break was forced via \\
}   % The line break was forced via \\}
% 1. For Chinese authors, the name in Chinese characters should also be given. For example, Gang Liu(刘刚), Xiao-Ming Li(李晓明)
% 2. Please ensure that every author approves the submission of the manuscript
% 3. Abbreviations should not be used in the affiliations
\date{\today}
\maketitle

\begin{abstract}
{
In this paper, we investigate the decoherence of qubit due to its coupling to a Hermitian or a non-Hermitian bath within the pure dephasing spin-boson model. First, using this model, we analytically establish the previously anticipated similarity between the non-equilibrium and the equilibrium correlation functions $P_x(t)$ and $C_x(t)$. Then, in the short/long time asymptotic behaviors of $P_x(t)$, we find singular dependence on $A$ (coupling strength) and $s$ (bath exponent) at their integer values. Finally, we find that the non-Hermitian bath tends to suppress the decoherence of qubit for all values of $A$ and $s$, in contrast to the conclusion of Dey {\it et al.}. Our results show the potential of non-Hermitian environment engineering in suppressing the decoherence of qubit. }

\end{abstract}


\section{Introduction}

The principle of superposition of states is one of the basements of quantum mechanics. At the macroscopic and mesoscopic scales, however, it is difficult to prepare and measure such superposition states because they tends to lose the superposition properties (such as interference effects) during the preparation and measurement. The research on this phenomenon is at the core of the fields such as quantum-to-classical transition, quantum computing and quantum information technology.\textsuperscript{\cite{Schlosshauer1}} The concept of quantum decoherence, as put forward by Zeh\textsuperscript{\cite{Zeh1}} contains the following main points. The quantum systems are never completely isolated from the environment in practice. When a quantum system interacts with the environment, it usually quickly entangles with a large number of environmental degrees of freedom, which will significantly affect the local observation results of the system. In particular, the quantum interference effect due to the state superposition will be effectively suppressed. It is thus extremely difficult to observe these physical quantities in most of the practical situations. The superposition state required for the quantum information processing is usually prone to decoherence. For this reason, decoherence has become a major obstacle to the realization of practical quantum information processing devices (such as quantum computers).\textsuperscript{\cite{Vion1}} 
	
In the past three decades, many methods have been proposed from the theoretical or experimental side to avoid or mitigate the decoherence. For examples, there are methods such as quantum error correction,\textsuperscript{\cite{Nielsen1}} decoherence-free subspace,\textsuperscript{\cite{Lidar1}} technique of dynamical decoupling,\textsuperscript{\cite{Viola1}} and the quantum Uncollapse.\textsuperscript{\cite{katz1}} Besides these methods, it is also proposed that the non-Hermitian effect can be employed to suppress the decoherence.\textsuperscript{\cite{Gardas1,Bhat1,Dey1,El-Ganainy1,Duttatreya1,Wang1,Li1,Li2}} The corresponding non-Hermitian system has been implemented in experiments.\textsuperscript{\cite{Ruter1,Gao1,Ashida1,Wu1}} All these studies not only deepen our understanding of the decoherence, but also improve the ability to artificially regulate the decoherence. The development of non-Hermitian physics also benefited from such studies.\textsuperscript{\cite{Ashida1}}
	
In quantum mechanics, the operator corresponding to the physical observable is a Hermitian operator. In particular, the Hamiltonian corresponding to the energy is a Hermitian operator. This is one of the most important assumptions of quantum mechanics, which ensures the probability conservation of the isolated quantum system and realness of energy. However, when we consider an open system, the energy, particles, and information exchange between the system and the environment.\textsuperscript{\cite{Ashida1}} Such a system can be described by a non-Hermitian Hamiltonian. Gamow\textsuperscript{\cite{Gamow1}} and Feshbach\textsuperscript{\cite{Feshbach1}} first used this idea to study the $\alpha$ decay problem, showing that the non-Hermitian Hamiltonian can play an important role in solving physics problems. As a milestone of the non-Hermitian physics development, Bender and Stefan found that a non-Hermitian Hamiltonian with the parity-time symmetry has real eigenvalues.\textsuperscript{\cite{El-Ganainy1}} Since then, many non-Hermitian Hamiltonian models with the parity-time symmetry have been realized in experiment.\textsuperscript{\cite{Ruter1,Gao1,Ashida1,Wu1}} Up to now, the non-Hermitian physics has gained significant progress both in theoretical and experimental aspects. 

Recently, some theoretical research has been carried out on the effect of non-Hermitian bath on the decoherence of qubit. There are two main types of research in this direction. In the first type of research, experimentally feasible non-Hermitian model Hamiltonians are proposed, in which the system and/or the environment part is non-Hermitian. By transforming the total Hamiltonian into an equivalent solvable Hamiltonian, the influence of non-Hermitian on quantum decoherence is studied. For example, Gardas\textsuperscript{\cite{Gardas1}} theoretically demonstrated that for the model of non-Hermitian spin coupled to the Hermitian boson system, for the Ohmic bath and the zero temperature limit, the non-Hermitian effect can suppress the decoherence of the uncorrelated initial state of the ground state. After that, Bhat\textsuperscript{\cite{Bhat1}} conducted a more in-depth analysis and found that for non-Ohmic bath and at finite temperature, the non-Hermitian effect can also suppress the decoherence of the uncorrelated and correlated initial states of the ground state. The corresponding non-Hermitian Hamiltonian has been experimentally realized recently,\textsuperscript{\cite{Ruter1,Gao1}} making the theoretical prediction testable in experiment. In 2019, Dey {\it et al.}\textsuperscript{\cite{Dey1}} constructed a non-Hermitian model Hamiltonian of the Hermitian system coupled to a non-Hermitian environment. By transforming it into an equivalent solvable model, they studied the influence of non-Hermitian environment on the quantum decoherence. However, in that work, the transformation is carried out for the system and environment degrees of freedom, but leave the coupling term intact. The result is therefore inconclusive. In 2025, Dey {\it et al.}\textsuperscript{\cite{Duttatreya1}} revealed that quantum coherence can be maximized when the system and the environment are simultaneously non-Hermitian and satisfy the PT symmetry condition. They proposed an experimental scheme based on the nitrogen-vacancy center in diamond and the optomechanical system for observing the effect. 

For the second type of research, the main steps of research are as follows: first, perform a priori non-Hermitian operation on the spin, then let the spin interact with the environment, and finally evaluate the fidelity of the initial state and the final state, so as to study the influence of non-Hermitian effect on the quantum entanglement and quantum decoherence.\textsuperscript{\cite{Wang1,Li1,Li2}}
	
This paper is organized as follows: In section 2, the model Hamiltonian is described. In section 3, we give the derivation and results for the decoherence factor, equilibrium correlation function, and the non-equilibrium correlation function. We discuss the relationship between them. In section 4 we present the decoherence for the non-Hermitian spin-boson model. The influence of the non-Hermitian bath on the decoherence of the spin is discussed and compared with the results of Dey {\it et al.}.; In section 5, we make a summary.

\section{Model}

In this work, we study the effect of non-Hermitian environment on the decoherence of a quantum two-level system at $T=0$ in the pure dephasing situation. Following the work of Dey {\it et al.}\textsuperscript{\cite{Dey1}}, we consider this problem using the pure dephasing spin-boson model with the non-Hermitian bath Hamiltonian,
\begin{equation}    \label{eq1}
  	  H^{NH} = H_{S}+H_{SB}+H_{B}^{NH},
\end{equation}
with
	\begin{eqnarray}   	  \label{eq2}
	  &&  H_{S} = \frac{\epsilon}{2}\sigma_{z},  \nonumber\\
	  && H_{SB}=\frac{\sigma_{z}}{2}\sum_{k}\lambda_{k} ( a_{k}^{\dagger} + a_{k} ),   \nonumber\\
  	  && H_{B}^{NH} = \sum_{k} \left( \frac{p_{k}^{2}}{2m}+\frac{1}{2}m\omega_{k}^{2}x_{k}^{2} + 2i\tau \omega_{k} p_{k}x_{k} \right).
	  \end{eqnarray}
Here, $\sigma_{z}$ is the Pauli matrix describing the  two-level system. $a_{k}$ is the annihilation operator of the environmental boson with energy $\omega_{k}$. The coupling strength $\lambda_{k}$ between spin and the $k$-th boson mode is real. In the bath Hamiltonian $H_{B}^{NH}$, $x_k$ and $p_k$ are the coordinate and momentum of bath mode $k$, respectively. The non-Hermiticity of the bath Hamiltonian is introduced by the third term, with $i$ being the imaginary number unit. The extent of non-Hermitian is characterized by the real and positive dimensionless parameter $\tau$. Due to the $\mathcal{PT}$ symmetry of $H_{B}^{NH}+H_{SB}$, the Hermiticity of $H_{S}$, and the commutativity between $H_{S}$ and $H_{B}^{NH}+H_{SB}$, the full Hamiltonian have real eigenvalues. Due to the absence of the tunnelling term $-\Delta \sigma_x$ in $H_{S}$, one can obtain the exact solution of spin dynamics for this Hamiltonian at certain initial conditions. Therefore, this Hamiltonian is a simple but nontrivial model suitable for investigating the pure dephasing effect.

To parametrize the bath, we first consider the limit $\tau =0$ in Eq. (\ref{eq1}). In this limit, the non-Hermitian Hamiltonian reduces to the standard Hermitian pure dephasing spin-boson model \textsuperscript{\cite{leggett1987dynamics,weiss2012quantum}}
	\begin{eqnarray}    \label{eq3}
		H_{SBM}=\frac{\epsilon}{2}\sigma_{z}+\sum_{k}\omega_{k}a_{k}^{\dagger}a_{k}+\frac{\sigma_{z}}{2}\sum_{k}\lambda_{k}(a_{k} + a_{k}^{\dagger}).	\nonumber\\ 
	\end{eqnarray}
The influence of the environment on the impurity is determined by the bath spectral function
	\begin{eqnarray}    		\label{eq4}
		J(\omega)\equiv \sum_{k}\pi\lambda_{k}^{2}\delta(\omega-\omega_{k}).
	\end{eqnarray}
In this work, we use the standard power-law form of the spectral function with an exponential high-frequency cut-off,
	\begin{eqnarray}   		\label{eq5}
		J(\omega)=\pi AB^{1-s}\omega^{s}e^{-\omega / B} \ \ \ \ (s>0).
	\end{eqnarray}
Here, $A$ is the dimensionless dissipative strength. $B$ is the the high-frequency cut-off energy scale of the spectral function. The bath is called sub-Ohmic, Ohmic, and super-Ohmic for the parameter regimes $0 < s <1$, $s=1$, and $s>1$, respectively. Eq. (\ref{eq1}) (with a tunneling term $-\Delta \sigma_x$ in general) can be realized in many physical systems and processes, such as superconducting circuits,\textsuperscript{\cite{Vion1}} quantum dots in solids,\textsuperscript{\cite{PBertet2005,PBorri2001}} cold atom systems,\textsuperscript{\cite{JKeeling2010,KBaumann2010}} and the energy transfer processes in photosynthesis.\textsuperscript{\cite{DXu1994,LAPachon2011}}

In the work of Dey {\it et al.} \textsuperscript{\cite{Dey1}}, Eq. (\ref{eq1}) is studied in the Ohmic regime by a similarity transformation method. The transformation was, however, only carried out for the Hamiltonians of the environment, but not for the coupling term between the system and environment. As a result, their conclusion applies only to the weak coupling limit. In this paper, we re-investigate this problem in a broader parameter regime that include also the sub-Ohmic and super-Ohmic baths. We also removed the defect in the original derivation of Dey {\it et al.} by the exact similarity transformation carried out for the full Hamiltonian. We obtain qualitatively different results from those of Dey. Dey shows that only for $s>1$ can the non-Hermitian bath suppress the decoherence in the long time limit. We find that the finite non-Hermiticity suppresses the decoherence of qubit for any time and $s$.

\section{Decoherence in the Hermitian spin-boson model}

By a series of transformations (see Section 4), the non-Hermitian pure dephasing model Eq. (\ref{eq1}) with $\tau >0$ can be reduced to an effective Hermitian model in the form of the standard spin-boson Hamiltonian. As a result, the dephasing behavior of the non-Hermitian system can be extracted from that of the Hermitian system. Therefore, in this section, we first presents results for the decoherence properties of the standard Hermitian spin-boson Hamiltonian. In the next section, we will discuss the non-Hermitian effect based on these results.

Below, we first review the well-known decoherence properties of the standard Hermitian model. Then we summarize our comparative study of the equilibrium and non-equilibrium time correlation function. All the study in this paper is for zero temperature.

\subsection{Decoherence function}

For the Hermitian model Eq. (\ref{eq1}) at $\tau=0$, Breuer obtained the integral expression of the decoherence function, for the initial state of uncorrelated system and bath.\textsuperscript{\cite{breuer2002theory}}
Suppose that at time $t=0$, the system and bath are uncorrelated, {\it i.e.},  $\rho_{SB}=\rho_{S}(0)\otimes \rho_{B}$. Here $\rho_{S}(0)$ is the density matrix of the system at time $t=0$ and $\rho_{B}$ is the density matrix of the bath in the thermal equilibrium state. The state of the total system at time $t >0$ reads $\rho_{SB}(t)=U_{SB}(t)\rho_{SB}U_{SB}^{+}(t)$, where $U_{SB}(t)$ is the time propagator of the total system. Tracing out the bath, we obtain 
\begin{eqnarray}\label{eq6}
		\rho_{S}(t) &=& Tr_{B}\left[\rho_{SB}(t) \right]  \nonumber \\
                    &=& 
	\begin{pmatrix}
		\rho_{S}^{00} & e^{-\gamma(t)}e^{-i\epsilon t}\rho_{S}^{01}  \\
		& \\
		e^{-\gamma(t)}e^{i\epsilon t}\rho_{S}^{10} & \rho_{S}^{11}
	\end{pmatrix},
\end{eqnarray}
with
\begin{eqnarray}    		\label{eq7}
		\gamma(t)=\int_{0}^{+\infty}\frac{J(\omega)}{\pi \omega^{2}} \left[1- \cos{(\omega t)} \right] \coth (\beta \omega/2 )  \, d\omega
\end{eqnarray}
being the decoherence factor. $\rho_{S}^{\alpha \beta}$ is the matrix element of $\rho_{S}(0)$ under the basis of the eigenstates of $\sigma_{z}$. $\beta =1/(k_B T)$ is the inverse temperature. Substituting Eq. (\ref{eq5}) into Eq. (\ref{eq7}) and taking $T=0$, Breuer obtained the decoherence function of the system in the Ohmic dissipative environment. Morozov \textsuperscript{\cite{morozov2012decoherence}} obtained the decoherence function of the system in the super-Ohmic and sub-Ohmic dissipative environments. These results are summarized as
\begin{equation}         \label{eq8}
\gamma(t)  =  \frac{1}{2}A\ln(1+B^{2}t^{2}),  \ \,\,\,\,\,\,\,\,\,\,\, (s=1)
\end{equation}
and
\begin{eqnarray}         \label{eq9}
 \gamma(t) &=& A\Gamma(s-1) \times \nonumber \\
&&   \left\{ 1-(1+B^2 t^{2})^{\frac{1-s}{2}}
\cos{\left[ (s-1) \arctan{(Bt)} \right] } \right\}  .   \nonumber \\
&&    \,\,\,\,\,\,\,\,\,\,\,\,\,\, \,\,\,\,\,\,\,\,\,\,\,\,\,\,  \,\,\,\,\,\,\,\,\,\,\,\,\,\,\,\,\,\,\,\,\,\,\,\,\,\,  \,\,\,\,\,\,\,\,\,\,\,\,\,\,\,\,\,\,\,\,\,\,\,\,\,\, \,\,\,\,\,\,\,\,\,\,\,  (s \neq 1)
\end{eqnarray}

\subsection{Equilibrium and  non-equilibrium correlation functions}

Here, we present our comparative study of the equilibrium and non-equilibrium correlation function.
Lü and Zheng \textsuperscript{\cite{lu2007quantum}} calculated the non-equilibrium correlation function $P_{\sigma_{z},\sigma_{z}}(t)$ of the full spin-boson model (with finite tunnelling term) to study the relaxation of the two-level system under the sub-Ohmic bath. They used the perturbative approach based on a unitary transformation. By comparing the numerical results, they found that $P_{\sigma_{z},\sigma_{z}}(t)$ has similar behavior with that of the equilibrium correlation function $C_{\sigma_{z},\sigma_{z}}(t)$ in small coupling and intermediate time. It is unknown yet to what extent this similarity is exact or universal.
Their definitions for $P_{A,B}(t)$ and $C_{A,B}(t)$ are as follows,
\begin{eqnarray}    	\label{eq10}
	{C}_{A,B}(t)=\frac{1}{2}\langle[A(0),B(t)]_{+}\rangle_{H}
\end{eqnarray}
\begin{eqnarray}        \label{eq11}
	{P}_{A,B}(t)=\frac{1}{2}\langle[A(0),B(t)]_{+}\rangle_{ini}.
\end{eqnarray}
Here, $[A,B]_{+} \equiv AB+BA$ is the anti-commutator of $A$ and $B$. $\langle\cdots \rangle_{H}$ stands for the average value in the thermodynamic equilibrium state of $H$. $\langle\cdots \rangle_{ini}$ stands for the average value in the given initial state.

  In this part, we present the exact analytic expressions of  $C_{x}(t) \equiv C_{\sigma_{x},\sigma_{x}}(t)$ and $P_x(t) \equiv P_{\sigma_{x},\sigma_{x}}(t)$ for the Hermitian pure dephasing model Eq. (\ref{eq1}) at $\tau=0$ and $T=0$, with the uncorrelated initial state for $P_{x}(t)$. These expressions will help us reveal, on the quantitative level, the possible relations between the two equilibrium and non-equilibrium correlation functions in the pure dephasing case. 
  
  Suppose we have a spin-bath uncorrelated initial state  $\rho_{SB}(0)=\rho_{S}(0)\otimes \rho_{B}$. We use the initial spin state $\rho_{s}(0) = |\rightarrow\rangle \langle\rightarrow|$, {\it i.e.}, the eigen-state of $\sigma_x$ with eigenvalue $+1$. $\rho_{B}=\exp(-\beta H_{B})/Tr_{B}\exp(-\beta H_{B})$ is the density operator of bath in the equilibrium state, and we use $|B\rangle$ to represent the equilibrium state of the bath. Here $\beta = 1/(k_B T)$ is the inverse temperature.

For this initial state, using Eq. (\ref{eq6}), we have at $T=0$
\begin{eqnarray}        	\label{eq12}
    	P_{x}(t) &=& \frac{1}{2}\langle[\sigma_{x}(0),\sigma_{x}(t)]_{+}\rangle_{ini}   \nonumber \\
    	&=& \langle \sigma_x(t) \rangle_{ini}   \nonumber \\
    	&=& e^{-\gamma(t)} \cos(\epsilon t).
\end{eqnarray}
In the above equation, the decoherence function $e^{-\gamma(t)}$ is given by Eqs. (\ref{eq8}) and (\ref{eq9}). It describes the decaying part of $P_{x}(t)$. The oscillating factor is dominated by the bias $\epsilon$. See Appendix A for a detailed derivation of Eq. (\ref{eq12}).

To calculate the equilibrium correlation function ${C}_{x}(t)=\frac{1}{2}\langle[\sigma_{x}(0),\sigma_{x}(t)]_{+}\rangle_{H}$, we introduce an unitary transformation $U=\exp(\sigma_{z}K)$, with $K=\sum_{k} \lambda_{k}(a_{k}^{\dagger}-a_{k})/2\omega_{k}$. Using this transformation, we transform $H_{SBM}$ of Eq. (\ref{eq1}) into a decoupled form,
\begin{eqnarray}
   	\tilde{H}_{SBM} &=& U H_{SBM} U^{-1} \nonumber\\
   	&=&-\sum_{k} \frac{\lambda_{k}^{2}}{4\omega_{k}}+\frac{\epsilon}{2}\sigma_{z}+\sum_{k}\omega_{k}a_{k}^{\dagger}a_{k}  \label{eq13}.
\end{eqnarray}
Defining  $\tilde{\sigma}_{x}(t) = U\sigma_{x}(t)U^{-1}$, the equilibrium correlation function 
\begin{equation}
    	C_{x}(t) = \frac{1}{2}\langle[\tilde{\sigma}_{x}(0),\tilde{\sigma}_{x}(t)]_{+}\rangle_{\tilde{H}_{SBM}}    \label{eq14}
\end{equation}	
is obtained at $T=0$ as
\begin{eqnarray}        	\label{eq15}
    	&& C_{x}(t)  \nonumber \\
    	&=& e^{-\gamma(t)} \cos(\epsilon t) \cos\left[ \int_{0}^{+\infty}\frac{J(\omega)}{\pi \omega^{2}} \sin(\omega t) \, d \omega \right]    \nonumber\\
    	&& + e^{-\gamma(t)} \sin(\epsilon t)  \sin\left[\int_{0}^{+\infty}\frac{J(\omega)}{\pi \omega^{2}} \sin(\omega t) \, d \omega\right] \text{sgn}(\epsilon). \nonumber \\
       &&    	
\end{eqnarray}	
See Appendix B for a detailed derivation of Eq. (\ref{eq15}). Note that Eq. (\ref{eq12}) and Eq. (\ref{eq15}) have the common factor of decoherence function $e^{-\gamma(t)}$. But their time dependence differ significantly at $\epsilon \neq 0$, where the second term in Eq. (\ref{eq15}) adds different oscillation. To simplify the problem, in the rest part of this paper, we only focus on the decoherence properties in the symmetric case $\epsilon =0$. At $\epsilon=0$, the two correlation functions are quite similar and we have
\begin{eqnarray}     \label{eq16}
C_{x}(t) = P_{x}(t)\phi(t),
\end{eqnarray}	
with $ \phi(t) =	\cos \left[ \int_{0}^{+\infty}\frac{J(\omega)}{\pi \omega^{2}} \sin(\omega t) \, d \omega \right]$.
The integral is given as
\begin{eqnarray}       	\label{eq17}
&& \int_{0}^{+\infty}\frac{J(\omega)}{\pi \omega^{2}} \sin(\omega t) \, d \omega 
  \nonumber \\
&&   \nonumber \\ 
&& =  \left\{\begin{array}{lll}
    	    A \arctan(Bt), \,\,\,\,\,\,\,\,\,\,\,\,\,\,\, (s=1)   \\
    		\\
    		A \Gamma(s-1)\left(1+B^2t^2 \right)^{\frac{1-s}{2}}
    		\sin\left[(s-1) \arctan(Bt) \right], \\
    		 \,\,\,\,\,\,\,\,  \,\,\,\,\,\,\,\, \,\,\,\,\,\,\,\, \,\,\,\,\,\,\,\, \,\,\,\,\,\,\,\, \,\,\,\,\,\,\,\, \,\,\,\,\,\,\,\,  \,\,\,\,\,\,\,\, \,\,\,\,\,\,\,\, \,\,\,\,\,\,\,\, \,\,\,\,\,\,\,\, \,\,\,\,\,\,\,\, \,\,\,\,\,\,\,\,(s \neq 1). 
    	\end{array} \right.  \nonumber \\
&&    	
\end{eqnarray}	

The long time asymptotic expression of $\phi(t)$ reads
\begin{eqnarray}	\label{eq18}
&& \phi (t \to \infty)  \nonumber\\
&&  \nonumber \\
&& = \left\{\begin{array}{lllllll}
 \cos[\frac{\pi A}{2}], &  \,\,\,  (s=1, \,\, A \neq \text{odd} ) \\
 \\
 (-1)^{(A-1)/2} \frac{A}{Bt},  &   \,\,\,  (s=1, \,\, A = \text{odd} ) \\ 
 \\	
 1,    &   \,\,\,  (s>1)  \\              
 \\
 \cos\left[ A\Gamma(s-1)  \cos ( \frac{\pi}{2} s) (Bt)^{1-s} \right],   &   \,\,\,  (s<1 ) .  \\
\end{array}\right.   \nonumber \\
&&    
\end{eqnarray}

\subsection{Asymptotic behavior in short and long time limit}

To further compare $C_{x}(t)$ and $P_{x}(t)$, in this part, we analyse the short time and long time asymptotic behaviors of them at $\epsilon=0$ and $T=0$. In the zero time limit $t \rightarrow 0$, $C_{x}( t \to 0) =P_{x}( t \to 0) = 1$ as expected from the definition of $C_{x}(t)$ and $P_{x}(t)$.
Detailed analysis of short/long time asymptotic behavior of $C_{x}(t)$ and $P_{x}(t)$ at $\epsilon=0$ and $T=0$ will be presented below for $s=1$ and $s \neq 1$ separately. These analyses not only quantifies the similarity of $C_x(t)$ and $P_x(t)$, but also reveal interesting singular behaviors of the asymptotic decoherence at integer values of the coupling strength $A$ and the bath exponent $s$, which have not been discussed in the literature.	
	
\subsubsection{\label{src:level2} Ohmic case: $s=1$}

At $s=1$, we have
\begin{eqnarray}      \nonumber     \label{eq19}
&& P_{x}(t) =  \left[1 + (Bt)^2 \right]^{-A/2},   \nonumber \\
&& C_{x}(t) = [1 + (Bt)^2]^{-A/2} \cos\left[A \arctan(Bt) \right].
\end{eqnarray}
This leads to the short time and long time asymptotic behaviors respectively as 
\begin{eqnarray}         \label{eq20}
P_{x}(t) \sim \left\{\begin{array}{lll}
1-\frac{1}{2} A(Bt)^{2},   &  \,\,\, (Bt \ll 1) \\
\\
(Bt)^{-A} , &  \,\,\, (Bt \gg 1). \\
\end{array}\right.   \nonumber \\
\end{eqnarray}
and
\begin{eqnarray}       \label{eq21}
C_{x}(t) \sim \left\{\begin{array}{lllll}
1-\frac{1}{2} A(1+A)(Bt)^{2},   &  \,\,\,(Bt \ll 1) \\
\\
\cos(\frac{\pi A}{2})(Bt)^{-A} , & \,\,\,(Bt \gg 1, \,\, A \ne \text{odd}) \\
\\
    (-1)^{\frac{A-1}{2}} A(Bt)^{-A-1},  & \,\,\, (Bt \gg 1, \,\, A = \text{odd}).
\end{array}\right.   \nonumber \\
&&  
\end{eqnarray}
When $A$ is close to but not equal to an odd integer, there is a crossover time scale $t_{cr}(A)$ separating the time domain into two different power-law regimes in Eq. (\ref{eq21}). The short time regime $t \ll t_{cr}$ is dominated by the $A= \text{odd}$ behavior and the long time regime  $t \gg t_{cr}$ obeys the power law of $A \neq \text{odd}$. We obtain $B t_{cr} = A |\tan(\pi A /2)|$, which diverges when $A$ approaches an odd integer.

%----------------------------------------------
\begin{figure}[h]       	\label{Fig1}
\begin{center}  %图片全局居中
\vspace{-0.5cm}
\includegraphics[width=520pt, height=360pt, angle=0]{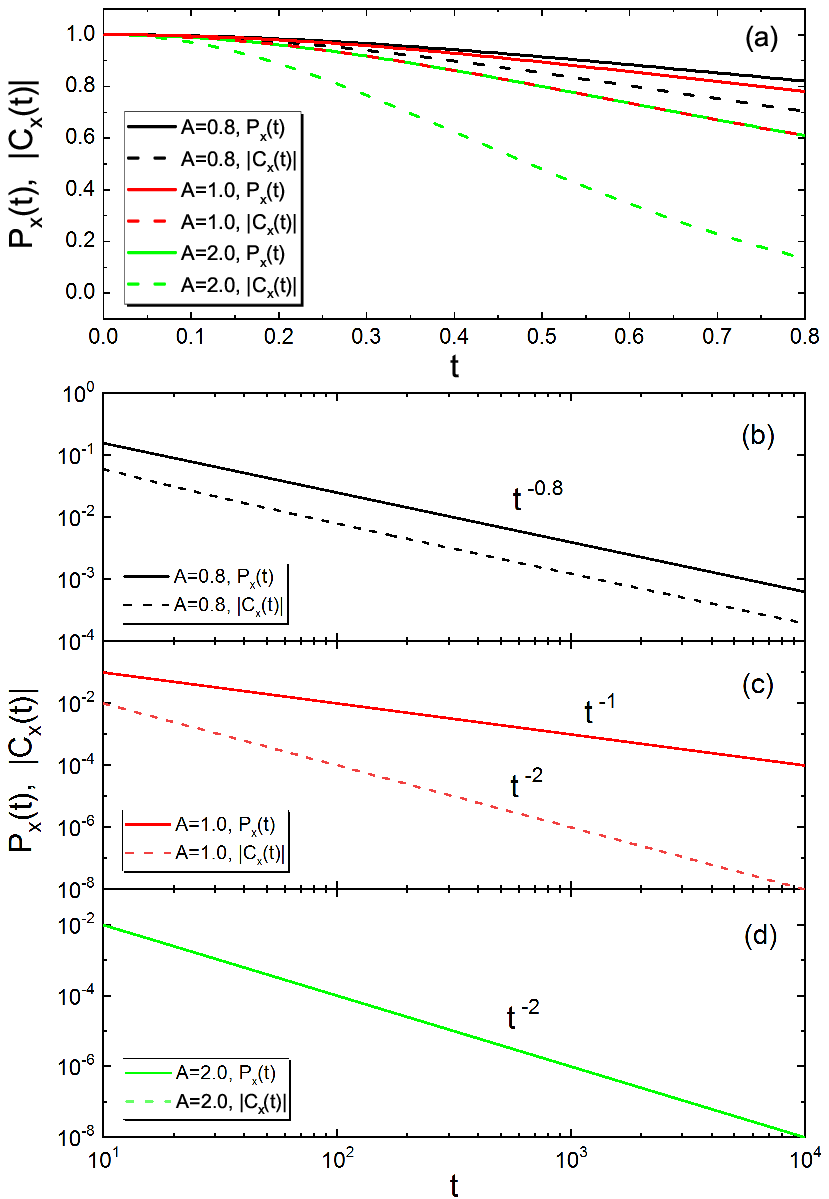}
\end{center}
\vspace*{-0.5cm}
\caption{(color online) $P_{x}(t)$ (solid line) and $|C_{x}(t)|$ (dashed line) as functions of time $t$ for $s=1$ and various $A$ values. (a) Short time regime; (b)-(d) long time regime. Parameters are $\epsilon=0$, $T=0$, and $B=1$. }
\end{figure}\label{Fig1}

These results are plotted and confirmed in Fig. 1 for $s=1.0$ and $B=1.0$. We compare $C_{x}(t)$ and $P_{x}(t)$ for three $A$ values, $A=0.8$, $1.0$, and $2.0$. Note that while $P_{x}(t) $ is always positive, $C_{x}(t)$ could become negative in the large $t$ limit due to the factor $\phi(t)$ in Eq. (\ref{eq18}). So we plot $|C_{x}(t)|$ for comparison. Fig. 1(a) is for the short time regime, while Figs.1(b) ($A=0.8$), (c) ($A=1.0$), and (d) ($A=2.0$) are for the long time regime. In Fig. 1(a), both $|C_{x}(t)|$ and $P_{x}(t)$ are quadratic function of $t$ in the small $t$ limit, with a factor $(1+A)$ difference in the $t^2$ term, as seen in Eqs. (\ref{eq20}) and (\ref{eq21}). Therefore, for small $t$,  $|C_{x}(t)|$ can be used to characterize the decoherence.

Figs. 1(b)-(d) show the power law decays of $|C_{x}(t)|$ and $P_{x}(t)$ in  the large $t$ regime. The have the same decay power $(Bt)^{-A}$ except for the odd integer $A$, {\it e.g.}, $A=1.0$ shown in Fig. 1(c). For odd integer $A$, $|C_{x}(t)| \sim (Bt)^{-A-1}$, decaying faster than $P_{x}(t) \sim (Bt)^{-A}$. 
Note that according to Eq. (\ref{eq19}), we have the exact relation $C_{x}(t)(A=1) = P_{x}(t)(A=2)$. This explains the coincidence of the red solid line with the green dotted line in Fig. 1(a).

\subsubsection{sub-Ohmic and super-Ohmic cases: $s \ne 1$}

For $s \ne 1$, we have $P_{x}(t) = e^{-\gamma(t)}$ and
\begin{equation}       	\label{eq22}
C_{x}(t) = e^{-\gamma(t)} \cos \left\{ A \Gamma(s-1) (1+B^2t^2)^{\frac{1-s}{2}} \sin \left[ (s-1) \arctan(Bt) \right] \right\}.   
\end{equation}
In the above equation,
\begin{equation} \label{eq23}
\gamma(t) = A\Gamma(s-1)\left\{ 1-[(1+(Bt)^{2}]^{\frac{1-s}{2}} \cos \left[ (s-1) \arctan(Bt) \right]  \right\}. 
\end{equation}
In the short time limit $Bt \ll 1$, the above expression gives
\begin{eqnarray}        	\label{eq24}
  &&  P_{x}(t)  \sim  1-\frac{1}{2}sA \Gamma(s) (Bt)^{2},    \nonumber \\
  &&  C_{x}(t) \sim     
        1-\frac{1}{2} A\Gamma(s) [A\Gamma(s)+s] (Bt)^{2}, \,\,\,\,\, (Bt \ll 1).  \nonumber \\
&&        
\end{eqnarray}
Again, they are quadratic function of $Bt$ and only the coefficients of $(Bt)^2$ term differ.

In the long time limit $Bt \gg 1$, we have the asymptotic behavior
\begin{eqnarray}     \label{eq25}
&& P_{x}(t) \sim 
\left\{\begin{array}{lll}
	P_0 e^{- \alpha_{1} (Bt)^{1-s}},  & \,\,\, (s\ne \text{even} )  \\
	\\
	P_0 e^{- \alpha_{2} (Bt)^{-s}}, & \,\,\, ( s = \text{even} ). 
\end{array}\right. 
\end{eqnarray}
Here $P_0 = e^{-A\Gamma(s-1)}$, $\alpha_1 = -A\Gamma(s-1) \sin(\frac{\pi}{2}s)$ and $\alpha_2 = A\Gamma(s)(-1)^{s/2}$. 
For the super-Ohmic bath with $s> 1$, $P_{x}(t)$ decays into a finite value in the large time limit $P_{x}(\infty) = P_0 > 0$ in a power law fashion $P_{x}(t)-P_x(\infty) \sim (Bt)^{1-s}$ ($s \neq \text{even}$ ) or $\sim (Bt)^{-s}$ ($s = \text{even}$). For $s$ close to but not equal to an even integer, the short time regime $Bt \ll Bt_{cr} = |\cot(\pi s/2)|/(s-1)$ is dominated by the even $s$ power law $P_{x}(t)-P_x(\infty) \sim (Bt)^{-s}$ while the long time regime $Bt \gg Bt_{cr}$ is dominated by the non-even $s$ power $P_{x}(t)-P_x(\infty) \sim (Bt)^{1-s}$. In the limit of $s$ approaching an even integer, $Bt_{cr}$ diverges.

For the sub-Ohmic case with $0 < s < 1$, $P_{x}(t)$ decays to zero exponentially fast with increasing $t$. Combined with the $s=1$ case where $P_x(t)$ decays in a power law $(Bt)^{-A}$ (Eq. (\ref{eq20})), we observe that the long time behavior of the decoherence function has a transition at $s=1$. This transition resembles an order-disorder phase transition of the equilibrium system in the thermodynamic limit, with $P_x(\infty)$ playing the role of an order parameter. $P_x(\infty) = 0$ in $0 < s < 1$, $P_x(\infty) =e^{-A \Gamma(s-1)} > 0$ in $s>1$, and $P_x(t)$ decays in a power law at $s=1$. This resemblance is not a coincidence. The spin-boson model can be mapped into a one-dimensional classical Ising model with long-range $S_z$-$S_z$ coupling $J_{ij} \sim |i-j|^{-(s+1)}$. \textsuperscript{\cite{weiss2012quantum}} In this mapping, the imaginary time plays the role of spatial distance between spins. For $0< s<1$, the long-range coupling of the Ising model gives rise to an ordered phase $\langle S_z \rangle >0$ even at finite temperature.\textsuperscript{\cite{Dyson1}} For $s>1$, the Ising model is in the disordered phase $\langle S_z \rangle =0$. $s=1$ is the critical range of coupling. When translated into the zero dimensional quantum model at zero temperature, these thermodynamic phase transitions correspond to the transitions in the behavior of dynamical correlation functions, with the correspondence $\langle S_z \rangle > 0 $ to $P_x(\infty) = 0$ in $0 < s< 1$, and $\langle S_z \rangle = 0 $ to $P_x(\infty) > 0$ in $ s > 1$.

The long time behavior of $C_{x}(t)$ is determined by both $P_x(t)$ and the oscillating function $\phi(t)$. We have the large $t$ limit of $\phi(t)$ as
\begin{eqnarray}   \label{eq26}
&& \phi(t \rightarrow \infty) = \nonumber \\
&& \left\{\begin{array}{lll}
\cos \left[A \Gamma(s-1)\cos(\frac{\pi}{2}s) (Bt)^{1-s} \right],  &  (s\ne \text{odd} )  \\
\\
\cos \left[ A \Gamma(s) (Bt)^{-s} \right], & ( s = \text{odd} ). 
\end{array}\right.   
\end{eqnarray}
For $s> 1$, $\phi(t)$ tends to unity in $t=\infty$. For $0 < s < 1$, $\phi(t)$ oscillates with decreasing frequencies in the large $t$ limit.
For $s$ being close to but not equal to an odd integer, the short time regime $Bt \ll Bt_{cr} = (s-1) |\tan(\frac{\pi}{2}s)|$ of $\phi(t)$ is dominated by the $s = \text{odd}$ behavior, and the long time regime $Bt \gg Bt_{cr}$ is dominated by the $s \neq \text{odd}$ behavior. $Bt_{cr}$ diverges as $s$ tends to an odd integer. Due to the non-linear argument $(Bt)^{1-s}$ or $(Bt)^{-s}$ in the cosin function, $\phi(t)$ has different behaviors in the long time limit. For $s>1$, $(Bt)^{1-s}$ and $(Bt)^{-s}$ decay to zero and $\phi(t)$ tends to unity without oscillation. For $0 < s < 1$, $(Bt)^{1-s}$ divergs in the large $t$ limit and $\phi(t)$ ocsillates with continually decreaing frequencies.

Combining $P_{x}(t)$ and $\phi(t)$ we obtain the asymptotic expression for $C_x(t)$ in $Bt \gg 1$ as
\begin{eqnarray}    \label{eq27}
 && C_{x}(t) \sim  \left\{\begin{array}{lllll}
P_0 e^{- \alpha_{1}(Bt)^{1-s}} \cos[\omega_{1}(Bt)^{1-s}], &   (s \ne \text{integer})
\\ 
\\
P_0 e^{- \alpha_{1}(Bt)^{1-s}} \cos[\omega_{2}(Bt)^{-s}],  & (s= \text{odd}) \\
\\
P_0 e^{- \alpha_{2}(Bt)^{-s}} \cos[\omega_{3}(Bt)^{1-s}],  & (s= \text{even}).
\end{array} \right.   \nonumber \\
&&
\end{eqnarray}
Here, $\alpha_1$ and $\alpha_2$ are given below Eq. (\ref{eq25}), and $\omega_i$ ($i=1,2,3$) are $\omega_{1} = A\Gamma(s-1) \cos(\frac{\pi}{2}s)$, $\omega_{2} = A\Gamma(s)$, and $\omega_{3} =A\Gamma(s-1)$. Similar to $P_x(t)$, in the limit $Bt = \infty$, $C_x(t)$ decays to zero exponentially fast for $0 < s < 1$. For $s>1$, it decays to a finite value $C_{x}(t=\infty) = P_0$, with $C_{x}(t)-C_x(\infty) \sim (Bt)^{1-s}$ ($s \neq \text{even}$ ) or $\sim (Bt)^{-s}$ ($s = \text{even}$), same as $P_x(t)-P_x(\infty)$. At $s=1$, it decays to zero in a power law (Eq. (\ref{eq21})). For $s>1$, there are special behaviors for $s$ being even and odd integers. For $s$ is close to but not equal to the integers, the short time regime $Bt \ll Bt_{cr}$ is dominated by the behavior of the integer $s$ and the long time regime $Bt \gg Bt_{cr}$ is dominated by the non-integer $s$ behavior. Here $Bt_{cr} = |\cot(\frac{\pi}{2}s)|/(s-1)$ and $Bt_{cr} = (s-1) |\tan(\frac{\pi}{2} s)|$ for even and odd $s$ values, respectively. In the limit $s = \text{integer}$, $Bt_{cr} = \infty$.

$C_x(t)$ differs from $P_x(t)$ by the additional cosin factor, as shown in Eq. (\ref{eq27}). As a result, apart from the decaying envelope, for $0 < s < 1$, $C_x(t)$ has an additional oscillation with continually increasing frequencies in the large $t$ limit. It is worth noting that it seems that the oscillation frequency increases with $t$ in Fig. 2(d), which is due to the selection of the logarithmic axis. In fact, the oscillation frequency is decreasing with t. In contrast, for $s \geq 1$, there is no oscillation in $C_x(t)$ in the long time limit.

%----------------------------------------------
\begin{figure} [H]      	\label{Fig2}
	\begin{center}  %图片全局居中
		\vspace{-0.5cm}
		\includegraphics[width=520pt, height=360pt, angle=0]{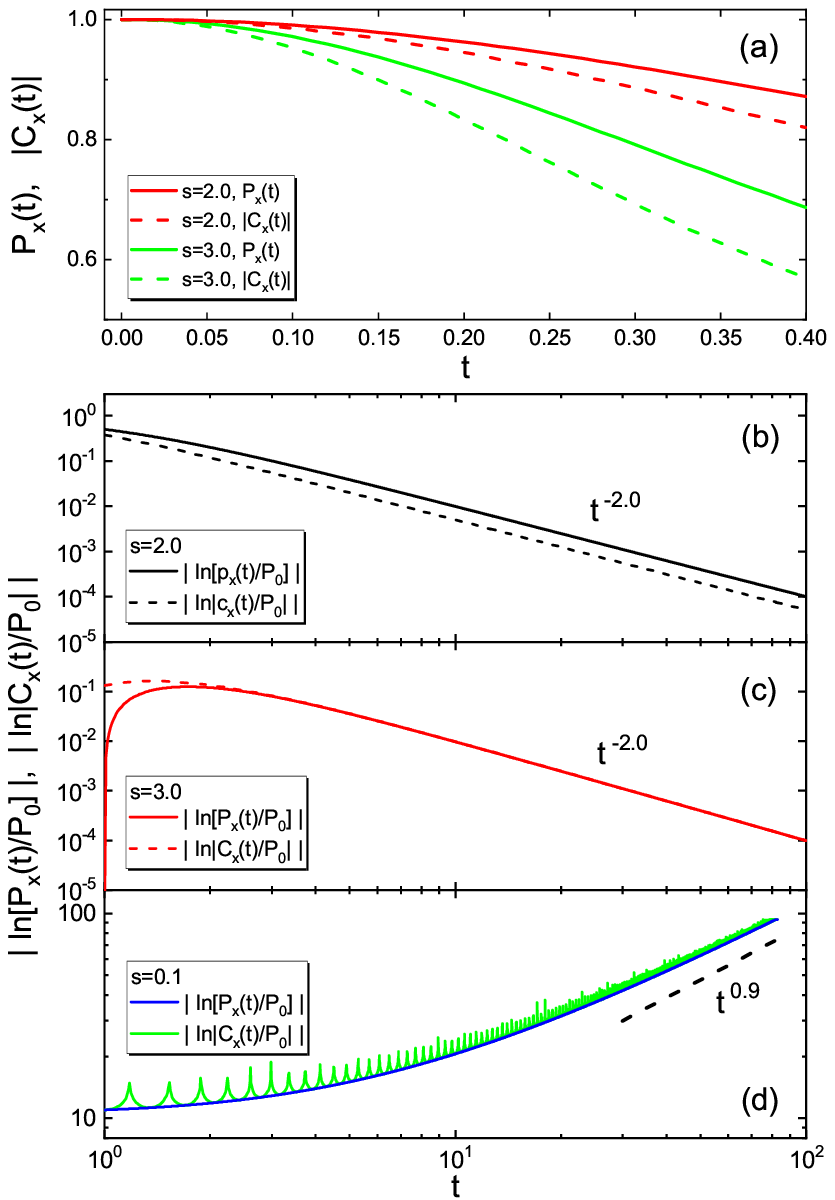}
	\end{center}
	\vspace*{-0.5cm} 
	\caption{(color online) (a) $P_x(t)$ and $|C_x(t)|$ for various $s$ values in the short time regime; (b)-(d): $|\ln[P_x(t)/P_0]|$ and $|\ln|C_x(t)/P_0||$ for various $s$ values in the long time regime. Parameters are $\epsilon=0$, $T=0$, $A=1$, and $B=1$. }
\end{figure}

    Fig. 2 shows the short/long time behaviors of $P_x(t)$ and $|C_x(t)|$ for the case $s \neq 1$. In Fig. 2(a), both $P_x(t)$ and $|C_{x}(t)|$ are quadratic functions in the short time regime, same as in the $s=1$ case. In Fig. 2(b)-(c), we plot $\ln[P_{x}(t)/P_0]$ and $|\ln| C_x(t)/ P_0||$ in the long time regime on the log-log scale. It exhibits the power law decay of $P_{x}(t)$ and $|C_x(t)|$ relative to their long time limits. That is, for $s > 1$, both $P_x(t)-P_0$ and $C_x(t) - P_0 $ decays as $(BT)^{\theta}$. For $0 < s < 1$, they decay in  the form $e^{-\alpha (BT)^{\theta}}$. Here the exponent $\theta$ is given by Eqs. (\ref{eq25}) and (\ref{eq27}). Note that $C_x(t)$ has an additional oscillation with decreasing frequency in the long time limit.

We conclude this section that for the unbiased pure dephasing spin-boson model, the long time behaviors of $C_{x}(t)$ and $P_{x}(t)$ at $T=0$ differ by a consin function of $t$. In the long time limit, the cosin function either tends to unity in the large $t$ limit (for $s \geq 1$) or oscillates with the frequency tending to zero (for $0 < s < 1$). The magnitudes of $C_x(t)$ and $P_x(t)$ have a common behavior in the large $t$ limit. They decay to zero exponentially fast in the form $e^{- \alpha t^{1-s}}$ ($\alpha >0$) for $0< s < 1$, to a finite value $P_0 > 0$ in a power law for $s>1$, or to zero in a power law for $s=1$. The exponent of the power law has special behavior at even and odd integer values of $s > 1$, or at odd integer values of $A$ for $s=1$.
Therefore, the equilibrium correlation function $C_{x}(t)$ indeed can be used to characterize the decoherence in the long time, except for the above stated special parameter points. 

\section{Decoherence in the Non-Hermitian spin-boson model}

Having obtained the detailed knowledge of the decoherence for the Hermitian spin-boson model, now we turn to the influence of the bath non-Hermiticity on the decoherence. We consider the spin-boson model Eq.(\ref{eq1}) with a non-Hermitian but $\mathcal{PT}$-symmetric bath, i.e., $\tau >0$.\textsuperscript{\cite{Dey1}}

\subsection{Effective Hermitian spin-boson model}

In the work of Ref.\cite{Dey1}, a similarity transformation $\eta= e^{\check{\tau}\sum_k p_{k}^2}$ ($\check{\tau} = \tau/ (m \omega_k \hbar)$) was applied to the bath Hamiltonian $H_B^{NH}$ of Eq.(\ref{eq2}) and transformed the latter into a Hermitian Hamiltonian $H_B^{H}$, being still quadratic in $a_k$ and $a_k^{\dagger}$. $H_B^{H}$ was then combined with the other two contributions $H_S + H_{SB}$ to form the starting point of the study. This practice does not lead to the exact decoherence of model $H_{NH}$ of Eq.(\ref{eq1}) since the change of $H_{SB}$ due to the transformation was not taken into account. This leads to the question as to what extent the conclusion of Ref.\cite{Dey1} hold in the full parameter space of the pure depahsing spin-boson model. 
	
In the present work, we will apply the exact similarity transformation to the {\it full } Hamiltonian $H^{NH}$ and change it into an effective Hermitian pure dephasing spin-boson model, $H_{u} = \eta H^{NH} \eta^{-1}$. Our results will hence be the exact ones for Eq.(\ref{eq1}). For this purpose, we introduce the transformation $\eta$ different from that of Dey as
	\begin{equation}      \label{eq28}
		\eta = e^{ - (m \omega_{k} \tau /\hbar) \sum_{k}x_{k}^{2}}.
	\end{equation}
This transformation will change the form of $H_{B}^{NH}$ but will leave both the coupling term $H_{SB}$ and the system part $H_s$ intact. It produces a Hermitian Hamiltonian 
	\begin{eqnarray} 	\label{eq29}
		H_{u} & = & H_{S} + H_{SB}    \nonumber \\
		& & + \sum_{k} \omega_{k} \left[a_{k}^{\dagger}a_{k}  
		+ \tau^{2} ( a_{k} + a_{k}^{\dagger} )^{2} + \tau + \frac{1}{2}  \right].
	\end{eqnarray}

The deformed bath Hamiltonian in the above equation can be diagonalized by the Bogoliubov transformation
	\begin{eqnarray}   \label{eq30}
		a_{k} = v_{k} b_{k} + u_{k} b_{k}^{\dagger} ,      \nonumber\\
		a_{k}^{\dagger} = u_{k}b_{k} + v_{k} b_{k}^{\dagger}.
	\end{eqnarray}
This leads to an effective standard pure depahsing spin-boson model with the renormalized parameters
\begin{equation}   \label{eq31}
		H_{u} = H_{S} + H_{B}^{eff} + H_{SB}^{eff},
\end{equation}
with
	\begin{eqnarray}     \label{eq32}
		H_{B}^{eff} & = & \sum_{k} \tilde{\omega}_{k} b_{k}^{\dagger} b_{k},    \nonumber\\
		H_{SB}^{eff} & = & \frac{1}{2}\sigma_{z} \sum_{k} \tilde{\lambda}_{k} \left( b_{k} + b_{k}^{\dagger} \right).
	\end{eqnarray}
The parameters of the new Hamiltonian read
\begin{eqnarray}     \label{eq33}
 \tilde{\omega}_{k} &=& \sqrt{1 + 4\tau^{2}} \, \omega_{k} ,   \nonumber \\ 
 \tilde{\lambda}_{k} &=& \left(1 + 4\tau^2 \right)^{-\frac{1}{4}} \lambda_{k}.
\end{eqnarray}
The bath now acquires a new spectral function
\begin{equation}    \label{eq34}
J({\omega}) = \pi \tilde{A} \tilde{B}^{1-s}{\omega}^{s} e^{-\omega/\tilde{B}},
\end{equation}
with $\tilde{A} = (1+4\tau^{2})^{-3/2} A$ and $\tilde{B} = (1+4\tau^{2})^{1/2} \, B $.
At $\tau=0$, Eq. (\ref{eq29}) reduces to Eq. (\ref{eq3}). The decoherence of qubit described by the the non-Hermitian spin-boson model can then be obtained by replacing $A$ and $B$ with $\tilde{A}$ and $\tilde{B}$, respectively, in the results of the standard pure depahsing model Eq. (\ref{eq3}).

\subsection{The influence of Non-Hermitianity  on decoherence}

Now we are in a position to analyse the influence of the bath non-Hermiticity on the decoherence. We characterize the decoherence of qubit with the non-equilibrium correlation function of Eq. (\ref{eq12}), $P_x(t) = e^{-\gamma(t)}$. From Eqs. (\ref{eq8}), (\ref{eq9}) and Eq. (\ref{eq33}), the decoherence factor $\tilde{\gamma}(t)$ of the non-Hermitian spin-boson model for $\epsilon=0$, $T=0$, and $\tau >0$ is obtained as
\begin{equation}         \label{eq35}
\tilde{\gamma}(t)  =  \frac{1}{2}\tilde{A}\ln(1+  \tilde{B}^{2}t^{2}),  \ \,\,\,\,\,\,\,\,\,\,\, (s=1)
\end{equation}
and
\begin{eqnarray}         \label{eq36}
\tilde{\gamma}(t) &=& \tilde{A} \Gamma(s-1) \times \nonumber \\
&&   \left\{ 1-(1+ \tilde{B}^2 t^{2})^{\frac{1-s}{2}}
\cos{\left[ (s-1) \arctan{(  \tilde{B}t)} \right] } \right\}  .   \nonumber \\
&&    \,\,\,\,\,\,\,\,\,\,\,\,\,\, \,\,\,\,\,\,\,\,\,\,\,\,\,\,  \,\,\,\,\,\,\,\,\,\,\,\,\,\,\,\,\,\,\,\,\,\,\,\,\,\,  \,\,\,\,\,\,\,\,\,\,\,\,\,\,\,\,\,\,\,\,\,\,\,\,\,\, \,\,\,\,\,\,\,\,\,\,\,  (s \neq 1)
\end{eqnarray}
Here, the renormalized parameters $\tilde{A}$ and $\tilde{B}$ read $\tilde{A} = (1+4\tau^{2})^{-3/2} A$ and $\tilde{B} = (1+4\tau^{2})^{1/2}B$.

%----------------------------------------------
\begin{figure} [h]      	\label{Fig3}
	\begin{center}  %图片全局居中
		\vspace{-4.7cm}
		\includegraphics[width=570pt, height=420pt, angle=0]{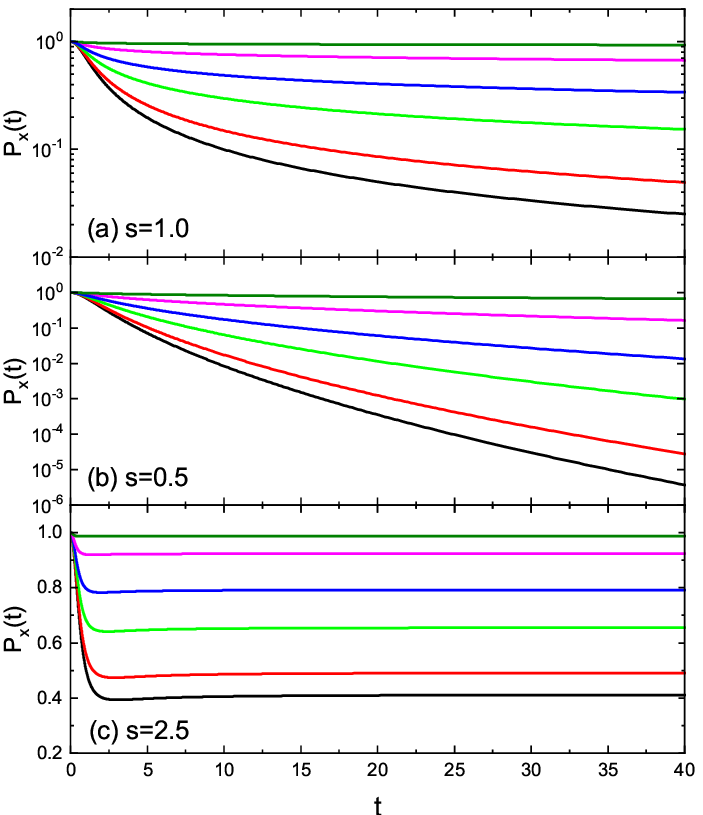}
	\end{center}
	\vspace*{-0.5cm} 
	\caption{(color online) $P_x(t)$ for various $s$ and $\tau$ values. (a) $s=1.0$, (b) $s=0.5$, and (c) $s=2.5$. In each panel, from bottom to top, $\tau=0.0$, $0.2$, $0.4$, $0.6$, $1.0$, and $2.0$. Parameters are $\epsilon=0$, $T=0$, $A=1$, and $B=1$. }
\end{figure}
%
%----------------------------------------------
\begin{figure} [h]      	\label{Fig4}
	\begin{center}  %图片全局居中
		\vspace{-4.9cm}
		\includegraphics[width=570pt, height=430pt, angle=0]{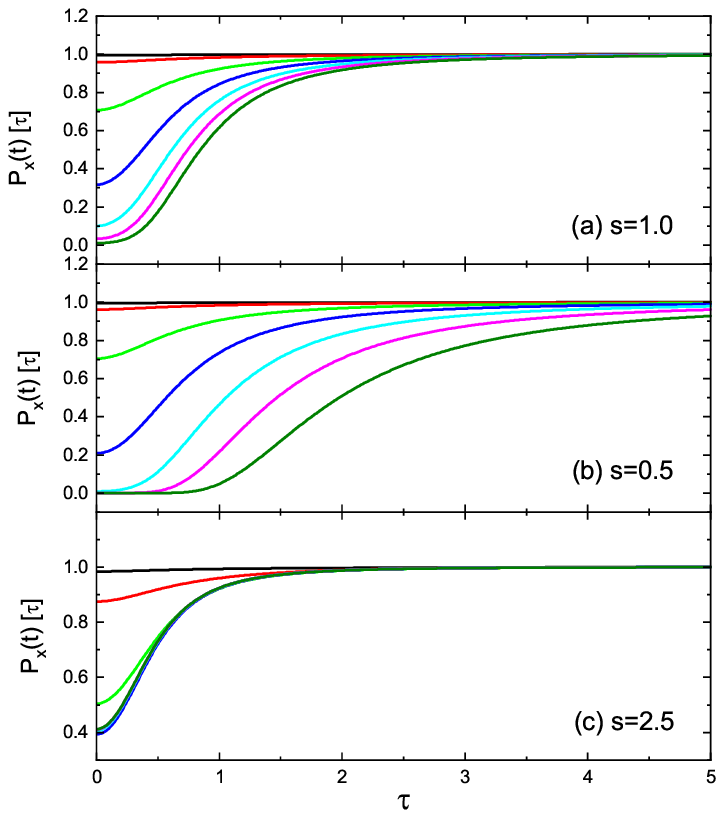}
	\end{center}
	\vspace*{-0.5cm} 
	\caption{(color online) $P_x(t)$ at fixed $t$ as functions of $\tau$. (a) $s=1.0$, (b) $s=0.5$, and (c) $s=2.5$. In each panel, from top to bottom, $t=0.1$, $0.3$, $1.0$, $3.0$, $10.0$, $30.0$, and $100.0$. Parameters are $\epsilon=0$, $T=0$, $A=1$, and $B=1$. }
\end{figure}

These expressions are different from those of Dey {\it et al.}. The results of Dey {\it et al.} do not have a simple renormalized form. The influence of the non-Hermitian bath on the qubit decoherence can thus be analyzed easily in our formalism in the full time regime.
$P_x(t) = e^{-\gamma(t)}$ is a decreasing function of $Bt$ for all $s$ and $A$ values. For $\tau >0$, $\tilde{Bt} > Bt$ and $\tilde{A} < A$. Replacing $Bt$ with $\tilde{B}t$ in the expression of $P_x(t)$ will shift the curve towards left and hence reduce $P_x(t)$. At the same time, replacing the factor $A$ by $\tilde{A}$ will increase it. The final effect depends on the competition between the two effects. Considering that both in the limits $Bt \ll 1$ and $Bt \gg 1$, $P_x(t)$ is a slowly varying function of $Bt$, we expect that the effect of $\tilde{A}$ will dominate the change and $P_x(t)$ will be increased by the non-Hermitian parameter $\tau > 0$. 

In Fig. 3, we present the generic curves of $P_x(t)$ at some $\tau$ values ranging from $\tau=0.0$ to $\tau = 2.0$. Fig. 3(a), (b), and (c) are for $s=1.0$, $0.5$, and $2.5$, representative for Ohmic, sub-Ohmic, and super-Ohmic baths. One can see that for different type of bath, although the initial decay speed and the long-time behavior of $P_x(t)$ differ significantly, all curves shift upward with increasing $\tau > 0$, showing that finite non-Hermiticity increases $P_x(t)$ and hence suppressed the decoherence.

A more quantitative analysis can be carried out by calculating $\partial P_x(t;\tau) / \partial \tau$. For $s=1$, we obtain the asymptotic results
 \begin{eqnarray}    \label{eq37}
\frac{\partial P_x(t;\tau)}{\partial \tau} &\approx&  \left\{\begin{array}{lll}
\frac{2}{1+4\tau^{2} } \tilde{A} (\tilde{B}t)^{2} \tau \,\,\, \geq  0,  &  ( Bt \ll 1 )  \\
 \\
\frac{12}{1+4\tau^{2} } (\tilde{B}t)^{-\tilde{A}} \tilde{A} \ln ( \tilde{B}t) \tau \,\,\, \geq 0,  & ( Bt \gg 1 ). 
\end{array}\right.   \nonumber \\
\end{eqnarray}
For $s \neq 1$, similar analytical calculation has been carried out and we find the same asymptotic behavior
 \begin{eqnarray}   \label{eq38}
 \frac{\partial P_x(t;\tau)}{\partial \tau} \geq 0, \,\,\,\,\,\, (Bt \ll 1  \,\,\, \,\,\, \text{or}  \,\,\, \,\,\, Bt \gg 1).
 \end{eqnarray}
The explicit expressions are too complicated and we will not present them here. 

For intermediate time, we examine the behavior by numerical calculation. In Fig. 4, we plot $P_x(t;\tau)$ at several fixed $t$ values as functions of $\tau$, for the same three $s$ values as in Fig. 3. Again, we find that for $t$ ranging from $0.1$ to $100.0$, $P_x(t;\tau)$ is an increasing function of $\tau$, supporting the conclusion that finite non-Hermiticity suppresses the decoherence of qubit.
 
The above conclusion is different from that of Dey {\it et al.}, which states that only for $s>1$ can the non-Hermitian bath suppress the decoherence in the long time limit. Our results therefore add to the significance of the possible utility of non-Hermiticity in reducing the decohrence of qubit in the experiment.

\section{Summary}

In this paper, we study the correlation functions of the pure dephasing SBMs with a Hermitian as well as with a non-Hermiticity bath. First, for SBM with Hermitian bath, we compare the rigorous expressions of the non-equilibrium correlation function $P_{x}(t)$ with the equilibrium one $C_{x}(t)$, establishing their formal similarity at $T=0$ and $\epsilon=0$. We then analyze their short- and long-time asymptotic behavior, disclosing the distinct forms at certain integer $A$ (for Ohmic case at $s=1$) and integer $s$ (for super- and sub-Ohmic case $s \neq 1$). Finally, the influence of a non-Hermitian bath on the decoherence is analyzed by transforming the latter to an effective Hermitian problem. We find that the finite non-Hermiticity parameter $\tau > 0$ suppresses the decoherence for general $A$ and $s$ values. These results may help people exploit the non-Hermiticity of environment to reduce the decoherenct of a qubit.

\addcontentsline{toc}{chapter}{Appendix A: Appendix section heading}

\section*{Appendix A: Derivation of non-equilibrium correlation function $P_{x}(t)$}

In this Appendix, we derive the expression of non-equilibrium correlation function $P_{x}(t)$, Eq. (\ref{eq12}). From the definition of $P_{x}(t)$, we have
\begin{eqnarray}        	\label{A1}
    	P_{x}(t) &=& \frac{1}{2}\langle[\sigma_{x}(0),\sigma_{x}(t)]_{+}\rangle_{ini}   \nonumber \\
    	&=&\frac{1}{2} \langle \sigma_{x}(0)\sigma_x(t)+\sigma_{x}(t)\sigma_{x}(0) \rangle_{ini}   \nonumber \\
     &=& \langle \rightarrow|  \langle B|  \sigma_x(t) |B \rangle | \rightarrow \rangle \nonumber \\
     &=& \langle \rightarrow|  \langle B|   e^{iHt}\sigma_x e^{-iHt} |B \rangle | \rightarrow \rangle\nonumber \\
     &=&\langle \sigma_x(t) \rangle_{ini}.  \nonumber \\
\end{eqnarray}
This gives the second equation of Eq.(\ref{eq12}). Further derivation gives
\begin{eqnarray}        	\label{A2}
    	P_{x}(t)&=&\langle \sigma_x(t) \rangle_{ini}   \nonumber \\
     &=& Tr_{S}\big\{ Tr_{B}[\rho_{SB}(t)]\sigma_{x} \big\}\nonumber \\
     &=& Tr_{S}[\rho_{S}(t)\sigma_{x}]\nonumber \\
     &=& e^{-\gamma(t)}e^{-i\epsilon t}\rho_{S}^{01}+e^{-\gamma(t)}e^{i\epsilon t}\rho_{S}^{10}.
\end{eqnarray}
For the initial spin state $| \rightarrow \rangle$,
$\rho_{S}^{01}=\rho_{S}^{10}=\frac{1}{2}$. Therefore, it can be obtained from Eq. (\ref{A2}) that
\begin{eqnarray}        	\label{A3}
    	P_{x}(t)&=&\frac{1}{2}[e^{-\gamma(t)}e^{-i\epsilon t}+e^{-\gamma(t)}e^{i\epsilon t}]\nonumber \\
             &=&e^{-\gamma(t)} \cos(\epsilon t).
\end{eqnarray}
This is the last equation of Eq.(\ref{eq12}).

\section*{Appendix B: Derivation of Eqs.(\ref{eq13}) and (\ref{eq15})}

In this Appendix, we derive the expressions of the transformed Hamiltonian Eq. (\ref{eq13}) and non-equilibrium correlation function $C_{x}(t)$, Eq. (\ref{eq15}).

For this purpose, below we will use the following Baker-Campbell-Hausdorff formulas
\begin{eqnarray}        	\label{B1}
    	e^{A}Be^{-A}=B+[A,B]+\frac{1}{2!}[A,[A,B]]+\frac{1}{3!}[A,[A,[A,B]]]... \, ,
\end{eqnarray}
and
\begin{eqnarray}        	\label{B4}
    	e^{A}e^{B}=e^{A+B}e^{[A,B]/2}, \ \ \ if\ [[A,B],A]=[[A,B],B]=0 .
\end{eqnarray}
The detailed proof of Eq. (\ref{B1}) and Eq. (\ref{B4}) can be found in Mehen 's note\textsuperscript{\cite{Mehen1}}.

For the unitary transformation $U=\exp(\sigma_{z}K)$, with $K=\sum_{k} \lambda_{k}(a_{k}^{\dagger}-a_{k})/2\omega_{k}$, it is easy to prove that
\begin{eqnarray}        	\label{B2}
  && Ua_{k}U^{-1} = a_{k}-\frac{\lambda_{k}}{2\omega_{k}}\sigma_{z},\nonumber \\
  &&  Ua_{k}^{\dagger}U^{-1} = a_{k}^{\dagger}-\frac{\lambda_{k}}{2\omega_{k}}\sigma_{z},\nonumber \\
 &&  U\sigma_{z}U^{-1} = \sigma_{z},\nonumber \\
 &&  \tilde{\sigma}_{x} = U\sigma_{x}U^{-1}= \sigma_{x}\cos(i2K)-\sigma_{y}\sin(i2K).
\end{eqnarray}
Here, $i$ is the imaginary unit.
From Eq. (\ref{B2}), we obtain
\begin{eqnarray}        	\label{B3}
    \tilde{H}_{SBM} &=& U H_{SBM} U^{-1} \nonumber\\
    &=&\frac{\epsilon}{2}\sigma_{z}+\sum_{k}\lambda_{k}\frac{\epsilon}{2}\sigma_{z}(a_{k}-\frac{\lambda_{k}}{\omega_{k}}\sigma_{z}+a_{k}^{\dagger})+\sum_{k}\omega_{k}(a_{k}-\frac{\lambda_{k}}{2\omega_{k}}\sigma_{z})(a_{k}^{\dagger}-\frac{\lambda_{k}}{2\omega_{k}}\sigma_{z})\nonumber\\
   	&=&-\sum_{k} \frac{\lambda_{k}^{2}}{4\omega_{k}}+\frac{\epsilon}{2}\sigma_{z}+\sum_{k}\omega_{k}a_{k}^{\dagger}a_{k},
\end{eqnarray}
which is Eq.(\ref{eq13}).

Next, we carry out the derivation of Eq. (\ref{eq15}).
For the harmonic oscillator Hamiltonian 
\begin{eqnarray}        	\label{B5}
H_{k}=\omega_{k}a_{k}^{\dagger}a_{k},
\end{eqnarray}        
there is the Bloch identity
\begin{eqnarray}        	\label{B6}
\langle e^{c}\rangle=e^{\frac{\langle c^{2} \rangle}{2}}, 
\end{eqnarray}        	
with $c=\lambda a_{k}^{\dagger}+\mu a_{k}$.
The detailed proof of Eq. (\ref{B6}) can be found in Gupta 's note\textsuperscript{\cite{Gupta1}}.
For the grand canonical ensemble, we therefore have
\begin{eqnarray}        	\label{B7}
\langle e^{c}\rangle=e^{\frac{\lambda \mu}{2}\frac{e^{\beta \omega_{k}}+1}{e^{\beta \omega_{k}}-1}}.
\end{eqnarray}        
We also need to use the following transformation of boson operators,
\begin{eqnarray}        	\label{B8}
e^{\lambda a_{k}^{\dagger}a_{k}}a_{k}e^{-\lambda a_{k}^{\dagger}a_{k}}&=&a_{k}e^{-\lambda},\nonumber\\
e^{\lambda a_{k}^{\dagger}a_{k}}a_{k}^{\dag}e^{-\lambda a_{k}^{\dagger}a_{k}}&=&a_{k}^{\dagger}e^{\lambda}.
\end{eqnarray}        	
Firs, we rewrite Eq. (\ref{eq13}) as
\begin{eqnarray}        	\label{B12}
\tilde{H}_{SBM}=H_{S}+H_{B}-\sum_{k}\frac{\lambda_{k}^{2}}{4\omega_{k}},
\end{eqnarray}
with $H_{S}=\frac{\epsilon}{2}\sigma_{z}$ and $H_{B}=\sum_{k}H_{k}=\sum_{k}\omega_{k}a_{k}^{\dagger}a_{k}.
$
In order to facilitate the following derivation, we rewrite $\tilde{\sigma}_{x}$ of Eq. (\ref{B2}) as
\begin{eqnarray}        	\label{B13}
\tilde{\sigma}_{x}=\frac{1}{2}(\sigma^{+}e^{-2K}+\sigma^{-}e^{2K})
\end{eqnarray}
with $\sigma^{+}=\sigma_{x}+i\sigma_{y}$ and $\sigma^{-}=\sigma_{x}-i\sigma_{y}$.

The expression for $C_x(t)$ becomes
\begin{eqnarray}    \label{B14}
    	C_{x}(t) &=& \frac{1}{2}\langle[\tilde{\sigma}_{x}(0),\tilde{\sigma}_{x}(t)]_{+}\rangle_{\tilde{H}_{SBM}} \nonumber\\  
                 &=&\frac{1}{8} \big[ \langle (\sigma^{+}e^{-2K})(\sigma^{+}e^{-2K})(t)\rangle_{\tilde{H}_{SBM}}+\langle (\sigma^{-}e^{2K})(\sigma^{+}e^{-2K})(t)\rangle_{\tilde{H}_{SBM}}\nonumber\\  
                      &&+\langle (\sigma^{+}e^{-2K})(\sigma^{-}e^{2K})(t)\rangle_{\tilde{H}_{SBM}}+\langle (\sigma^{-}e^{2K})(\sigma^{-}e^{2K})(t)\rangle_{\tilde{H}_{SBM}}\nonumber\\ 
                      &&+\langle (\sigma^{+}e^{-2K})(t)(\sigma^{+}e^{-2K})\rangle_{\tilde{H}_{SBM}}+\langle (\sigma^{-}e^{2K})(t)(\sigma^{+}e^{-2K})\rangle_{\tilde{H}_{SBM}}\nonumber\\
                      &&+\langle (\sigma^{+}e^{-2K})(t)(\sigma^{-}e^{2K})\rangle_{\tilde{H}_{SBM}}+\langle (\sigma^{-}e^{2K})(t)(\sigma^{-}e^{2K})\rangle_{\tilde{H}_{SBM}} \big]  \nonumber\\
                &=&\frac{1}{8} \big[ \langle \sigma^{+}e^{iH_{S}t}\sigma^{+}e^{-iH_{S}t}\rangle_{H_{S}}\langle e^{-2K}e^{iH_{B}t}e^{-2K}e^{-iH_{B}t}\rangle_{H_{B}}+\langle \sigma^{-}e^{iH_{S}t}\sigma^{+}e^{-iH_{S}t}\rangle_{H_{S}}\langle e^{2K}e^{iH_{B}t}e^{-2K}e^{-iH_{B}t}\rangle_{H_{B}}\nonumber\\
&&+\langle \sigma^{+}e^{iH_{S}t}\sigma^{-}e^{-iH_{S}t}\rangle_{H_{S}}\langle e^{-2K}e^{iH_{B}t}e^{-2K}e^{-iH_{B}t}\rangle_{H_{B}}+\langle \sigma^{-}e^{iH_{S}t}\sigma^{-}e^{-iH_{S}t}\rangle_{H_{S}}\langle e^{2K}e^{iH_{B}t}e^{2K}e^{-iH_{B}t}\rangle_{H_{B}}\nonumber\\
&&+\langle e^{iH_{S}t}\sigma^{+}e^{-iH_{S}t}\sigma^{+}\rangle_{H_{S}}\langle e^{iH_{B}t}e^{-2K}e^{-iH_{B}t} e^{-2K}\rangle_{H_{B}}+\langle e^{iH_{S}t}\sigma^{-}e^{-iH_{S}t}\sigma^{+}\rangle_{H_{S}}\langle e^{iH_{B}t}e^{2K}e^{-iH_{B}t} e^{-2K}\rangle_{H_{B}}\nonumber\\
&&+\langle e^{iH_{S}t}\sigma^{+}e^{-iH_{S}t}\sigma^{-}\rangle_{H_{S}}\langle e^{iH_{B}t}e^{-2K}e^{-iH_{B}t} e^{2K}\rangle_{H_{B}}+\langle e^{iH_{S}t}\sigma^{-}e^{-iH_{S}t}\sigma^{-}\rangle_{H_{S}}\langle e^{iH_{B}t}e^{2K}e^{-iH_{B}t} e^{2K}\rangle_{H_{B}} \big]. \nonumber\\
  &&
 \end{eqnarray}
Now we analyze all the eight terms in the above equation. Due to the time translation symmetry of the thermal equilibrium state, the last four terms can be obtained by simply changing the $t$ of the first four terms into $-t$. So we only need to consider the calculation of the first four terms. Because $[\sigma_{z},H_{S}]=0$, so both the first and the fourth terms are zero. So we just need to calculate the second and third terms.

It is easy to obtain
\begin{eqnarray}       \label{B15}
  \langle \sigma^{-}e^{iH_{S}t}\sigma^{+}e^{-iH_{S}t}\rangle_{H_{S}}=\frac{4e^{-\frac{\beta \epsilon}{2}}e^{i\epsilon t}}{e^{-\frac{\beta \epsilon}{2}}+e^{\frac{\beta \epsilon}{2}}}     \nonumber\\
  \langle \sigma^{+}e^{iH_{S}t}\sigma^{-}e^{-iH_{S}t}\rangle_{H_{S}}=\frac{4e^{\frac{\beta \epsilon}{2}}e^{-i\epsilon t}}{e^{-\frac{\beta \epsilon}{2}}+e^{\frac{\beta \epsilon}{2}}} .
\end{eqnarray}
Here, $\beta=\frac{1}{k_{B}T}$.
Since the bosons with different k indices are independent of each other, we have
\begin{eqnarray}        \label{B16}
 \langle e^{2K}e^{iH_{B}t}e^{-2K}e^{-iH_{B}t}\rangle_{H_{B}}&=&\prod \limits_{k}\langle e^{\frac{\lambda_{k}}{\omega_{k}}(a_{k}^{\dagger}-a_{k})}e^{i\omega_{k}ta_{k}^{\dagger}a_{k}}e^{-\frac{\lambda_{k}}{\omega_{k}}(a_{k}^{\dagger}-a_{k})}e^{-i\omega_{k}ta_{k}^{\dagger}a_{k}}\rangle_{H_{k}}\nonumber\\
  &=&\prod \limits_{k}\langle e^{\frac{\lambda_{k}}{\omega_{k}}(a_{k}^{\dagger}-a_{k})}e^{\frac{\lambda_{k}}{\omega_{k}}(e^{-i\omega_{k}t}a_{k}-e^{i\omega_{k}t}a_{k}^{\dagger})}\rangle_{H_{k}}\nonumber\\
  &=&\prod \limits_{k}e^{i\frac{\lambda_{k}^{2}}{\omega_{k}^{2}}\sin{(\omega_{k}t)}} \langle e^{\frac{\lambda_{k}}{\omega_{k}}a_{k}^{\dagger}(1-e^{i\omega_{k}t})-\frac{\lambda_{k}}{\omega_{k}}a_{k}(1-e^{-i\omega_{k}t})}\rangle_{H_{k}}\nonumber\\
  &=&\exp \left\{\sum_{k}i\frac{\lambda_{k}^{2}}{\omega_{k}^{2}}\sin{(\omega_{k}t)} -\sum_{k}\frac{\lambda_{k}^{2}}{\omega_{k}^{2}} \left[1-\cos{ (\omega_{k}t) } \right]\frac{e^{\beta \omega_{k}}+1}{e^{\beta \omega_{k}}-1} \right\}.
\end{eqnarray}
Eq. (\ref{B8}) is used in the derivation from the first line to the second line of Eq. (\ref{B16}). Eq. (\ref{B4}) is used in the derivation from the second line to the third line of Eq. (\ref{B16}). Eq. (\ref{B6}) is used in the derivation from the third line to the fourth line of Eq. (\ref{B16}).

Since Eq. (\ref{B16}) remains unchanged under transformation $\lambda_{k}\rightarrow-\lambda_{k}$, we have
\begin{eqnarray}     \label{B17}
 \langle e^{-2K}e^{iH_{B}t}e^{2K}e^{-iH_{B}t}\rangle_{H_{B}}&=& \langle e^{2K}e^{iH_{B}t}e^{-2K}e^{-iH_{B}t}\rangle_{H_{B}}\nonumber\\
&=&\exp \left\{\sum_{k}i\frac{\lambda_{k}^{2}}{\omega_{k}^{2}}\sin{ (\omega_{k}t) }-\sum_{k}\frac{\lambda_{k}^{2}}{\omega_{k}^{2}} \left[ 1-\cos{ (\omega_{k}t) } \right]\frac{e^{\beta \omega_{k}}+1}{e^{\beta \omega_{k}}-1}   \right\}.
\end{eqnarray}
Using Eq. (\ref{B15}), Eq. (\ref{B16}), and Eq. (\ref{B17}), we obtain
\begin{eqnarray}       \label{B18}
  	C_{x}(t) &=& \cos(\epsilon t) \cos \Big[ \sum_{k}\frac{\lambda_{k}^{2}}{\omega_{k}^{2}}\sin(\omega_{k}t) \Big] \exp \left\{ -\sum_{k}\frac{\lambda_{k}^{2}}{\omega_{k}^{2}} \big[ 1-\cos(\omega_{k}t) \big] \coth(\beta\omega_{k}/2)  \right\}     \nonumber\\
&& + \sin(\epsilon t) \sin\big[ \sum_{k}\frac{\lambda_{k}^{2}}{\omega_{k}^{2}}\sin(\omega_{k}t) \big] \tanh \left( \frac{\epsilon}{2k_{B}T} \right) \exp\left\{ -\sum_{k}\frac{\lambda_{k}^{2}}{\omega_{k}^{2}} \big[1-\cos(\omega_{k}t) \big] \coth(\beta\omega_{k}/2) \right\}. \nonumber\\  
&&
\end{eqnarray}
Finally, from Eq. (\ref{eq4}) and Eq. (\ref{eq7}), we obtain
\begin{eqnarray}    \label{B19}
  	C_{x}(t) &=& e^{-\gamma(t)}\cos(\epsilon t)\cos\left[ \int_{0}^{+\infty}\frac{J(\omega)}{\pi \omega^{2}} \sin(\omega t) \, d \omega \right] \nonumber\\
&& + e^{-\gamma(t)} \sin(\epsilon t) \tanh\left( \frac{\epsilon}{2k_{B}T} \right) \sin \left[\int_{0}^{+\infty}\frac{J(\omega)}{\pi \omega^{2}} \sin(\omega t) \, d \omega \right]. \nonumber\\  
&&
\end{eqnarray}
In the limit $T \rightarrow 0$, we get Eq. (\ref{eq15}).

\addcontentsline{toc}{chapter}{Acknowledgment}

\section*{Acknowledgment}
We acknowledge B. Miao for helpful discussions. This work is supported by NSFC (Grant No.11974420).

%\end{CJK*}  %% end the Chinese environment
%\end{document}  %%% end document
\newpage


\addcontentsline{toc}{chapter}{References}
\begin{thebibliography}{99}\footnotesize
\itemsep=-3pt plus.2pt minus.2pt   % set the reference line spacing
\bibitem{Schlosshauer1} Schlosshauer M \href{https://doi.org/10.1016/j.physrep.2019.10.001}{2019 \emph{Phys. Rep.} \textbf{831} 1} %1

\bibitem{Zeh1} Zeh H D \href{https://link.springer.com/article/10.1007/BF00708656}{1970 \textit{Found. Phys.} \textbf{1} 69}               %2
	
\bibitem{Vion1} Vion D, Aassime A, Cottet A, Joyez P, Pothier H, Urbina C, Esteve D, and Devoret M H  \href{https://www.science.org/doi/full/10.1126/science.1069372} {2002 \textit{Sci.Rep.} \textbf{296} 886} %3
		
\bibitem{Nielsen1} Nielsen M A and Chuang I L \href{https://www.cambridge.org/highereducation/books/quantum-computation-and-quantum-information/01E10196D0A682A6AEFFEA52D53BE9AE#overview} {2010 \textit{Quantum Computation and Quantum Information} 10th Anniversary edition (Cambridge: Cambridge University Press) pp.425—493} %4
		
\bibitem{Lidar1} Lidar D A, Bacon D, and Whaley K B \href{https://doi.org/10.1103/PhysRevLett.82.4556} {1998 \textit{Phys. Rev. Lett.} \textbf{82} 4556}%5
		
\bibitem{Viola1} Viola L, Knill E, and Lloyd S \href{https://doi.org/10.1103/PhysRevLett.82.2417} {1999 \textit{Phys. Rev. Lett.} \textbf{82} 2417}%6
		
\bibitem{katz1} Katz N, Neeley M, Ansmann M,Bialczak R C, Hofheinz M, Lucero E, O’Connell A, Wang H, Cleland A N, Martinis J M, and Korotkov A N \href{https://doi.org/10.1103/PhysRevLett.101.200401} {2008 \textit{Phys. Rev. Lett.} \textbf{101} 200401} %7

\bibitem{Gardas1} Gardas B, Deffner S, and Saxena A \href{https://doi.org/10.1103/PhysRevA.94.040101} {2016 \textit{Phys. Rev. A} \textbf{94} 040101}%8

\bibitem{Bhat1} Bhat J M, Lone M Q, Datta S, Dar G N, and Farouk A \href{https://www.academia.edu/83756873/Decoherence_in_a_PT_symmetric_qubit} {2023 \textit{Ukr. J. Phys.} \textbf{68} 101}%9

\bibitem{Dey1} Dey S, Raj A, and Goyal S K \href{https://doi.org/10.1016/j.physleta.2019.125931} {2019 \textit{Phys. Lett. A} \textbf{383} 125931} %10

\bibitem{El-Ganainy1} El-Ganainy R, Makris K G, Khajavikhan M, Musslimani Z H, Rotter S, and Christodoulides D N \href{https://www.nature.com/articles/nphys4323} {2018 \textit{Nature Phys.} \textbf{14} 11} %11

\bibitem{Duttatreya1} Duttatreya, Mohanty S, and Dey S \href{https://doi.org/10.1016/j.aop.2025.170298} {2026 \textit{Ann. Phys} textbf{484} 170298} %12

\bibitem{Wang1} Wang Q, Xu L, and He Z \href{https://iopscience.iop.org/article/10.1088/1555-6611/abc613} {2020 \textit{Laser Phys.} \textbf{30} 125203} %13

\bibitem{Li1}Li J X, Xu L, Zhao Y H, He Z, and Wang Q \href{https://iopscience.iop.org/article/10.1088/1555-6611/ac089d} {2021 \textit{Laser Phys.} \textbf{31} 075202} %14

\bibitem{Li2}Li J X, Chang S L, Zhao Y H, Xiao X and Leng Y \href{https://iopscience.iop.org/article/10.1088/1555-6611/ac1600} {2021 \textit{Laser Phys.} \textbf{31} 095202}%15

\bibitem{Ruter1} Ruter C E, Makris K G, El-Ganainy R, Christodoulides D N, Segev M, and Kip D \href{https://www.nature.com/articles/nphys1515} {2010 \textit{Nat. Phys.} \textbf{6} 192}%16

\bibitem{Gao1} Gao T, Estrecho E, Bliokh K Y, Liew T C H, Fraser M D, Brodbeck S, Kamp M, Schneider C, Hofling S, Yamamoto Y, Nori F, Kivshar Y S, Truscott A, Dall R, and Ostrovskaya E A \href{https://www.nature.com/articles/nature15522} {2015 \textit{Nature(London)} \textbf{526} 554}%17

\bibitem{Ashida1} Ashida Y, Gong Z P, and Ueda M \href{https://doi.org/10.1080/00018732.2021.1876991} {2020 \textit{Adv. Phys.} \textbf{69} 249}%18

\bibitem{Wu1} Wu Y, Liu W, Geng J, Song X, Ye X, Duan C K, Rong X, and Du J \href{https://www.science.org/doi/full/10.1126/science.aaw8205} {2019, \textit{Science} \textbf{364} 878}%19

\bibitem{Gamow1} Gamow G \href{https://link.springer.com/article/10.1007/bf01343196} {1928 \textit{Z. Physik.} \textbf{51} 204}%20

\bibitem{Feshbach1} Feshbach H, Porter C E, and Weisskopf V F \href{https://doi.org/10.1103/PhysRev.96.448} {1954 \textit{Phys. Rev.} \textbf{96} 448}%21

\bibitem{leggett1987dynamics} Leggett A, Chakravarty S, Dorsey A T, Fisher M P A, Grag A, and Zwerger W \href{https://doi.org/10.1103/RevModPhys.59.1} {1987 \textit{Rev. Mod. Phys.} \textbf{59} 1} %22

\bibitem{weiss2012quantum} Weiss U \href{https://www.worldscientific.com/worldscibooks/10.1142/6738#t=aboutBook} {2008 \textit{Quantum dissipative systems} Third Edition (London: World Scientific) pp.31-38, 293-294}%23
			
\bibitem{PBertet2005} Bertet P, Chiorescu I, Burkard G, Semba K, Harmans C J P M, DiVincenzo D P, and Mooij J E \href{https://doi.org/10.1103/PhysRevLett.95.257002} {2005 \textit{Phys. Rev. Lett.} \textbf{95} 257002} %24
			
\bibitem{PBorri2001} Borri P, Langbein W, Schneider S, and Woggon U \href{https://doi.org/10.1103/PhysRevLett.87.157401} {2001 \textit{Phys. Rev. Lett.} \textbf{87} 157401}%25
			
\bibitem{JKeeling2010} Keeling J, Bhaseen M J, and Simons B D \href{https://doi.org/10.1103/PhysRevLett.105.043001} {2010 \textit{Phys. Rev. Lett.} \textbf{105} 043001}%26
			
\bibitem{KBaumann2010} Baumann K, Guerlin C, Brennecke F, and Esslinger T \href{https://www.nature.com/articles/nature09009} {2010 \textit{Nature} \textbf{464} 1301}%27
			
\bibitem{DXu1994} Xu D and Schulten K \href{https://doi.org/10.1016/0301-0104(94)00016-6} {1994 \textit{Chem. Phys.} \textbf{182} 91} %28
			
\bibitem{LAPachon2011} Pachon L A and Nrumer P \href{https://pubs.acs.org/doi/10.1021/jz201189p} {2011 \textit{J. Phys. Chem. Lett.} \textbf{2} 2728}%29

\bibitem{breuer2002theory} Breuer H P and Petruccione F \href{https://academic.oup.com/book/27757?login=true} {2002 \textit{The theory of open quantum systems} First Edition (Oxford: Oxford University Press)}%30

\bibitem{morozov2012decoherence} Morozov V G, Mathey S, and R{\"o}pke G \href{https://doi.org/10.1103/PhysRevA.85.022101} {2012 \textit{Phys. Rev. A} \textbf{85} 022101}%31

\bibitem{lu2007quantum} L{\"u} Z G and Zheng H \href{https://doi.org/10.1103/PhysRevB.75.054302} {2007 \textit{Phys. Rev. B} \textbf{75} 054302}%32

\bibitem{Dyson1} Dyson F J \href{https://link.springer.com/article/10.1007/BF01645907} {1969 \textit{Commun. Math. Phys.} \textbf{12} 91}%33

\bibitem{Mehen1} Mehen T \href{https://webhome.phy.duke.edu/~mehen/760/ProblemSets/BCH.pdf} {Notes on Baker-Campbell-Hausdorff (BCH) Formulae Duke University}

\bibitem{Gupta1} Gupta K D \href{https://homepages.iitb.ac.in/~kdasgupta/pdf/ThermalAverage.pdf} {Why is “Thermal Average” needed how is it done? Indian Institute of Technology Bombay}
\end{thebibliography}
\end{document}